\documentclass[10pt,letterpaper]{article}
\usepackage[top=0.85in,left=2.75in,footskip=0.75in]{geometry}
\usepackage{stfloats}
\usepackage{changepage}
\usepackage{amsmath}

\usepackage[utf8x]{inputenc}

 \usepackage{booktabs}

\usepackage{textcomp,marvosym}

\usepackage{lipsum,tabularx}
\usepackage{amsmath,amssymb}

\usepackage{graphicx}
\usepackage[justification=centering]{subfig}
\usepackage[colorlinks]{hyperref}

\usepackage{cite}

\usepackage{nameref,hyperref}

\usepackage[right]{lineno}
\usepackage{color}
\usepackage{microtype}
\DisableLigatures[f]{encoding = *, family = * }

\raggedright
\usepackage[aboveskip=1pt,labelfont=bf,labelsep=period,justification=raggedright,singlelinecheck=off]{caption}

\makeatletter
\renewcommand{\@biblabel}[1]{\quad#1.}
\makeatother

\date{}

\usepackage{lastpage,fancyhdr,graphicx}
\usepackage{epstopdf}
\fancyheadoffset[L]{2.25in}
\fancyfootoffset[L]{2.25in}

\begin{document}
\vspace*{0.2in}

\begin{flushleft}
{\Large
\textbf\newline{Modelling Lobbying Behaviour and Interdisciplinarity Dynamics in Academia} % Please use "title case" (capitalize all terms in the title except conjunctions, prepositions, and articles).
}
\newline
\\
Stefano Mazzoleni\textsuperscript{1+}, %\Yinyang},
%\Yinyang},
Lucia Russo\textsuperscript{2}, %\Yinyang},
Francesco Giannino\textsuperscript{1}, %\Yinyang},
Gerardo Toraldo\textsuperscript{3}, %\Yinyang},
Constantinos I. Siettos\textsuperscript{4*} 

%Name4 Surname\textsuperscript{2},
%Name5 Surname\textsuperscript{2\ddag},
%Name6 Surname\textsuperscript{2\ddag},
%Name7 Surname\textsuperscript{1,2,3*},
%with the Lorem Ipsum Consortium\textsuperscript{\textpilcrow}
\bigskip
\textbf{1}  Laboratory of Applied Ecology and System Dynamics, Dipartimento di Agraria, University of Naples Federico II, Italy
\\
\textbf{2} Consiglio Nazionale delle Ricerche, Naples, Italy
\\
\textbf{3} Dipartimento di Matematica e Applicazioni ``Renato Caccioppoli", University of Naples Federico II, Italy
\\
\textbf{4} School of Applied Mathematics and Physical Sciences, National Technical University of Athens, Greece\

+stefano.mazzoleni@unina.it, *ksiet@mail.ntua.gr

\end{flushleft}
\section*{Abstract}
 
Disciplinary diversity is being recognized today as the key to establish a vibrant academic environment with bigger potential for breakthroughs in research and technology. However, the interaction of  several factors including policies, and behavioral attitudes put significant barriers on advancing interdisciplinarity. A ``cognitive rigidity" may rise due to reactive academic lobbying favouring inbreeding. Here, we address, analyse and discuss a mathematical model of lobbying and interdisciplinarity dynamics in Academia. The model consists of four coupled non-linear Ordinary Differential Equations simulating the interaction between three types of academic individuals and a state reflecting the rate of knowledge advancement which is related to the level of disciplinary diversity. Our model predicts a rich nonlinear behaviour including multiplicity of states and sustained periodic oscillations resembling the everlasting struggle between the ``new" and the ``old". The effect of a control policy that inhibits lobbying is also studied. By appropriate adjustment of the model parameters we approximated the jump/phase transitions in breakthroughs in mathematical and molecular biological sciences resulted by the increased flow of Russian scientists in the USA after the dissolution of the Soviet Union starting in 1989, the launch of the Human Genome Project in 1992 and the Internet diffusion starting in 2000. 

\linenumbers

\section*{Introduction}

The challenging complex problems that we are facing today, the ones with pressing important social, health and environmental impacts (such as the climate change, the (re)emergence and the spread of infectious diseases and the mapping of the human brain connectome) are beyond the potential of any single scientific discipline to confront by itself. Their solution requires the synergy and integration of knowledge and efforts from diverse scientific disciplines. Thus, the role and importance of disciplinary diversity and its efficient integration is recognized today as a key to unlocking the potential to achieve breakthroughs in science and technology \cite{Holl,Rafols,Aboelela,Wonseok,Misra,Silvius,NAS,BritishAcademy}. 
Recall the example of computational neuroscience, a prototype of interdisciplinary subject that was born around the early 70s fertilized by the pioneering work of Ian Hodgkin and Andrew Huxley (Nobel Prize laureates in Medicine in 1963). The need to understand the complexity of brain development and functioning led to the integration of different traditional disciplines ranging from medicine and biology to physical, social sciences, mathematics and computer science.\\
Over the last decades, interdisciplinarity has been emerged by two ways \cite{Weingart}: (a) internally through the interaction of different disciplines themselves in the mill of science (such as the case of the birth of computational neuroscience \cite{Schutter}), and (b) externally, driven by political decisions that allocate public science funding (such as in the case of the Human Genome Project). \\
 Today, many Universities, government agencies and Institutions acknowledge the importance to foster multi- and interdisciplinary interactions both in research and teaching programs \cite{Silvius,NAS,BritishAcademy,Aboelela}. However, this process is neither monotonic nor easy to establish. Structural, behavioural and conflict of interests, especially in funding, raise significant barriers \cite{Nissani,Pellmar,BritishAcademy}. All in all, what is recognized as the main barrier %to the advance of diversity and interdisciplinarity in established academic envirnonments 
is the resistance to change \cite{Ledford}. These barriers establish a ``cognitive rigidity’ that favors the conduction of both research and teaching within the rigid boundaries of disciplines \cite{Pellmar}. Policies and practices for the allocation of research grants, recruitment, tenure and promotion are some of the structural barriers hindering interdisciplinarity \cite{Nissani,BritishAcademy,Lattuca}. This structural ``rigidity" coupled with professional friendships/lobbying make established/reactive practices harder to change \cite{BritishAcademy}. As in any social system, people are connected to others that share common and more familiar to them practices for better or for worse. High rates of academic inbreeding in the same disciplines are identified in many academic systems around the world making recruitment completely impermeable to external candidates \cite{Horta,Bleike,Navaro}.
In the past, this practice  was likely to be beneficial in terms of fast production of research results as it fosters research team cohesion and continuity and diminishes risks recruitment\cite{Horta}. However, the challenges of today demand openness and disciplinary diversity to innovation \cite{BritishAcademy,NAS}.\\
Here, we address a mechanistic dynamical model in the form of coupled nonlinear ordinary differential equations (ODEs) that aspires to simulate the interplay between lobbying practices and disciplinary diversity at the level of a department.

\begin{figure*}[!ht]
\begin{center}
\includegraphics[scale=0.7]{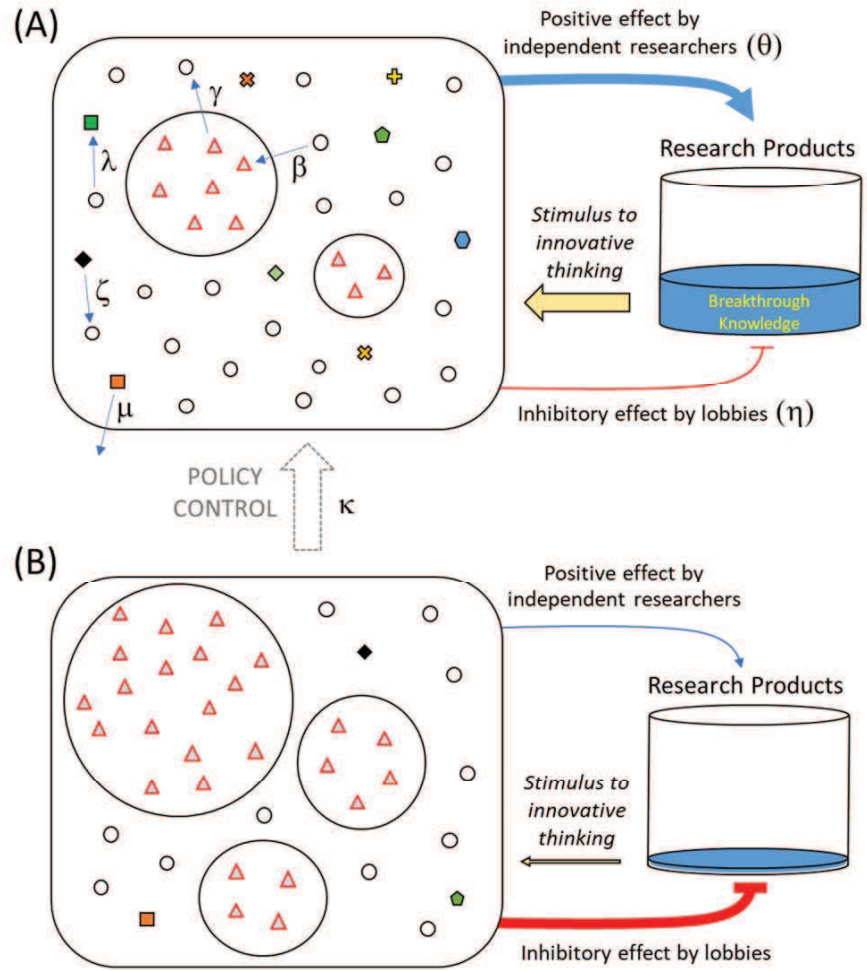}
\\
\caption{Conceptual overview of the model. Schematic representation of academic system structure and related effects on rates of breakthrough knowledge advancement. Black filled symbols refer to independent researchers in different fields, triangles and open circles represent lobby and neutral individuals, respectively. In (A) the system is characterized by high disciplinary diversity and small sized lobbies, thus maintaining sustainable conditions for positive rates of knowledge advancement and consequent strong stimulus to innovation. In (B) the system presents large sized lobbies and therefore low discipline diversity, producing a negative feedback on the rate of knowledge ``jumps" and consequent reduced stimulus to breakthroughs. The transition from B to A is only possible by an external policy control against lobbying. Greek letters refer to model parameters see Table 1}
\label{fig1}
\end{center}
\end{figure*}

Our model (see Figure \ref{fig1} for a conceptual scheme) contains three different types of individuals:  Lobbyists, Independent researchers and Neutrals. Their interactions and relative dominance create the scientific-cultural environment affecting the capability of the system to achieve breakthroughs in knowledge and innovation.
Lobbyists mainly act strongly defending their research field/discipline, thus favouring academic inbreeding. Their target is to increase their academic power (rising the number of researchers in their group and career advancement for the group members). Possibly they may also try to create a ``friendly'' network between different groups, thus creating ``cartels” that share similar goals in relation with fund raising strategies. Hence, lobbyists show a clear tendency to work more within their sectorial expertise on mainstream topics rather than on new unexplored ideas. On the other hand, independent researchers are individuals more interested in research questions and knowledge advancement than in academic career and management. Persons belonging to this category have a genuine interest and tendency to work across traditional disciplinary boundaries. In fact, this attitude has been explicitly recognized as a transdisciplinary orientation characterizing researchers with higher production of interdisciplinary research articles \cite{Misra}. 
Thus, lobbyists put barriers and obstacles to the careers of independent researchers by influencing the recruitment process in a biased way to the advantage of their own growth.  However, we are not stating that it is intrinsically wrong to try to favour the own group, as this can be the natural case also for an independent researcher. The main difference in our perspective relies in the level of hostility against the others. An independent researcher appreciates a good active group, independently of its affiliation. Differently, a lobbyist perceives as a risk the growth of a different group. These two opposite behaviours objectively produce feedbacks on the diversity levels of the academic scenario with lobbyists and independent researchers decreasing and increasing, respectively in the long term, the rate of knowledge innovation. 
Moreover, Neutral individuals represent those researchers that do not provide strong contribution to knowledge breakthroughs, but at the same time do not actively participate to lobby groups. However, these individuals may change their status either joining a lobby group or becoming an independent researcher. Such a decision is influenced  in an analogy to the concept of ``cultural attractors''  by two opposing factors: the ``power" of lobbyists and the level of attractiveness of interdisciplinarity, respectively \cite{Claidi,Buskell}. \\
We also address a fourth variable (the potential for breakthrough knowledge, $BK$), representing the effect of the level of knowledge integration of disciplinary diversity. By definition $BK$ is enhanced by independent researchers and inhibited by lobbyists because disruptive technologies and/or breakthrough concepts most likely rise from cross-border interactions rather than from data accumulation within an established field \cite{BritishAcademy,NAS}.\\

The model dynamics have been systematically studied with the aid of numerical bifurcation theory to explore the solutions in the parametric space and detect the critical points that mark the onset of phase changes in the academic structure and the related capability of innovation. Within this context, the bifurcation diagrams were constructed in the one and two dimensional parametric space with respect to the external policy intensity aiming at establishing disciplinary diversity and the ``power" of lobbyists to recruit/inbreed neutrals. The model exhibits a complex dynamical behaviour including multistability of equilibria, stable (and unstable) sustained oscillations and turning points that mark the boundaries between lobbying prevalence and corresponding establishment of disciplinary diversity or maintenance of a disciplinary ``silo" and successive feedbacks on knowledge advancement. Thus, by calibration of the values of the model parameters corresponding to the rates of advancement/inhibition of inderdisciplinarity, the intensity of control policy against lobbying and the transition rates between neutrals and individuals, we simulated the phase transitions in interdisciplinary research in biology reported from 1989 to 2000 due to the flow of ex-soviet scientists in the USA, the Human Genome Project  and the Internet diffusion.

%~\cite{saito,porter,izhik,izhik2}) and experiments~\cite{tzortzakis}.
%

\section*{Results}

\begin{figure*}[!ht]
\begin{center}
\includegraphics[scale=0.72]{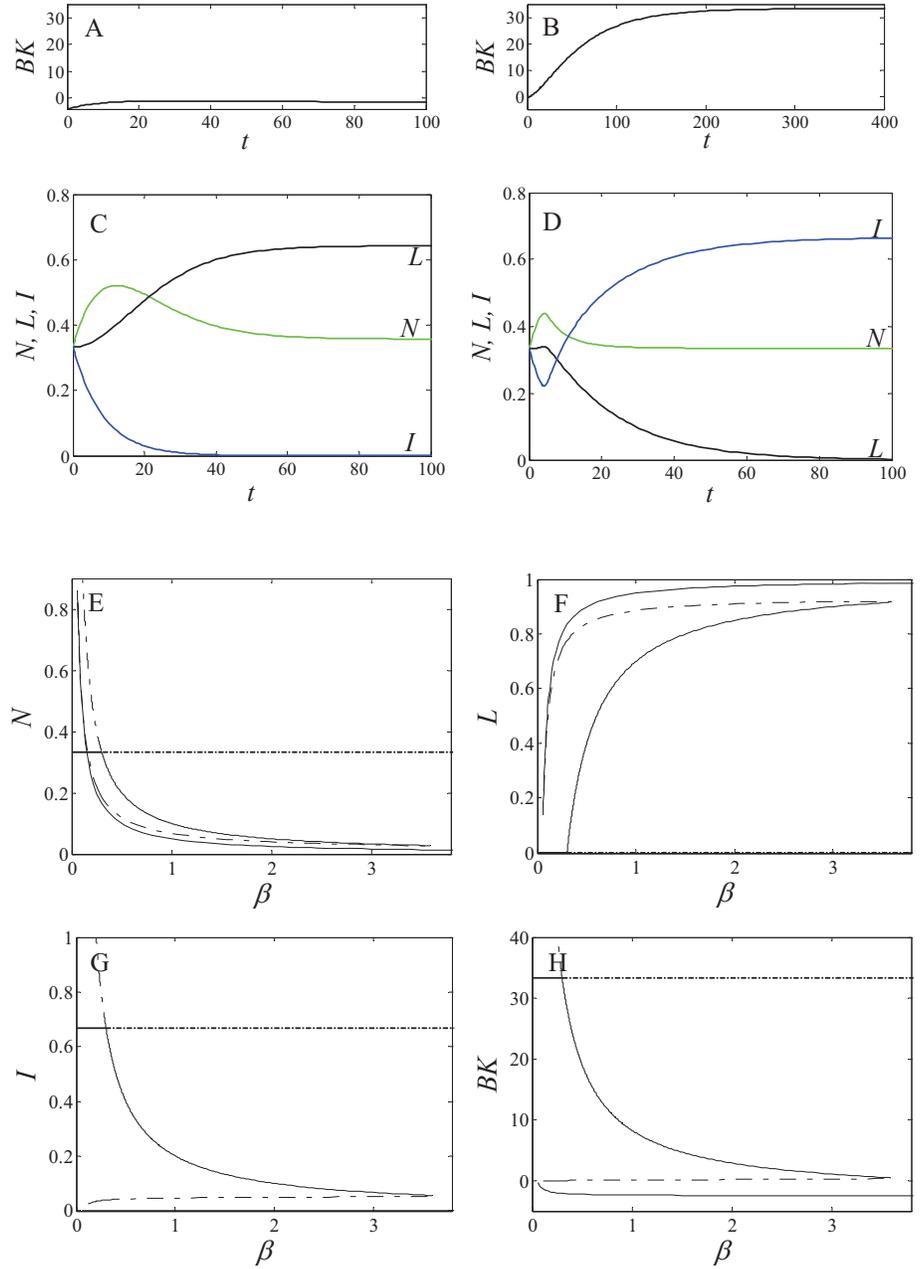}
\\
\caption{\textbf{Model dynamics without control policy ($\kappa=0$, $\beta=0.14$ )}. \textbf{Time evolution of the state variables}. \textbf{(A)} $BK$ and \textbf{(C)} $N, L, I$ with initial conditions $N(0)=0.334$, $L(0)=0.333$, $I(0)=0.333$, $BK(0)=-4$, \textbf{(B)} $BK$ and \textbf{(D)} $N, L, I$ with initial conditions $N(0)=0.334$, $L(0)=0.333$, $I(0)=0.333$,  $BK(0)=-1$.\\  \textbf{One-dimensional bifurcation diagrams w.r.t. $\beta$}. \textbf{E}. $N$, \textbf{F}. $L$, \textbf{G}. $I$, \textbf{H}. There are two turning points, $TP_1$ at $\beta \approx 0.058$ ($(N, L, I, BK) = (0.862,0.136,0.001,-0.284)$), and, $TP_2$ at $\beta  \approx 3.588$ ($(N, L, I, BK) = (0.028,0.917,0.054,0.428)$); a transcritical bifurcation ($TR$) appears also at $\beta \approx 0.3$ ($(N, L, I, BK) = (0.333, 0, 0.667,33.333)$). Solid lines correspond to stable equilbria and dashed lines to unstable equilbria.}
\label{fig2}
\end{center}
\end{figure*}

The system of ODEs  Eq.~(\ref{eqn11}-\ref{eqn14}) (See Materials and Methods) was analysed using the tools of numerical bifurcation analysis. 
We analyzed the model dynamics with respect to the intensity of the power of lobbyists to recruit neutrals (represented by the parameter $\beta$) and the intensity of the external policy  to tackle with lobbying behaviour (represented by the parameter $\kappa$). The values of the other parameters are $\mu=0.5$, $\lambda=0.1$, $\gamma=0.05$, $\zeta=0.07$, $\theta=1$, $\eta=0.05$, $\epsilon=0.02$.\\
In the absence of control policy, i.e. for $\kappa=0$, and for $\beta=0.14$, the system exhibits multiplicity of states. Depending on the initial conditions of the states $(N(0), L(0), I(0), BK(0))$, two different stable equilibrium points can be reached (see Figure \ref{fig2}A-D). Starting for example with equal densities for all three types of academic behaviour and $BK(0)=-4$, the system converges to $(N, L, I, BK) \approx (0.357, 0.643, 0, -1.185)$. This state is characterized by a low level of breakthrough knowledge capability, dominated by the lobbying behaviour and zero number of independent researchers (Figure \ref{fig2}A,C). Setting as initial condition $BK(0)=-1$ the system converges to a lobby-free state $(N, L, I, BK) \approx (0.333, 0, 0.667, 33.333)$ characterized by a very high level of breakthrough knowledge potential and dominance of the independent  researchers (Figure \ref{fig2}B,D).

To systematically investigate the system's behaviour w.r.t.$\beta$, in the absence of a control policy, we constructed the bifurcation diagrams shown in Figure \ref{fig2}E-H. 

There are two turning points at $\beta \approx 0.058$ ($(N, L, I, BK)$ $\approx (0.862,0.136,0.001,-0.284)$) ($TP_{1}$) and at $\beta \approx 3.588$ ($(N, L, I, BK)$ $\approx (0.028,0.917,0.054,0.428)$) ($TP_{2}$), as well as a transcritical bifurcation ($TR$) at $\beta \approx 0.3$ ($(N, L, I, BK) \approx (0.333, 0, 0.667,33.333)$). The two turning points mark the appearance and disappearance of solutions and the transcritical bifurcation marks the exchange of the stability between the solution branches. Thus, for $\beta<0.058$, the only stable solution is the lobby-free state characterized by the dominance of Independent researchers and therefore very high levels of potential for breakthroughs. At this branch, a transcritical bifurcation ($TR$) appears at which the lobby-free state looses its stability in favour of another branch of stable equilibria that ends up to the turning point $TP_{2}$. This branch is now characterized by the co-existence of all three ``species" and thus by relatively high levels of disciplinary diversity. For example on this branch at $\beta \approx 0.5$, $(N, L, I, BK) \approx (0.2,0.4,0.4,19)$, while at $\beta \approx 1.0$, $(N, L, I, BK) \approx (0.1,0.7,0.2,8.313)$. Beyond $TP_{1}$, the solutions become unstable up to the other turning point ($TP_{1}$). At $TP_{1}$ a branch of stable equilibria is born. This branch is characterized by the absence of independent researchers ($I=0$) (see Figure \ref{fig2}G) and thus low levels of potential of breakthroughs (see Figure \ref{fig2}H). For increasing values of $beta$, this branch leads asymptotically to the absolute dominance of lobbyists. Indeed, on this branch at $\beta \approx 0.125$, $(N, L, I, BK) \approx (0.4,0.6,0,-1.5)$, while for $\beta \approx 0.5$, $(N, L, I, BK) \approx (0.9,0.1,0,-2.25)$. \\

\begin{figure*}[!ht]
\begin{center}
\includegraphics[scale=0.65]{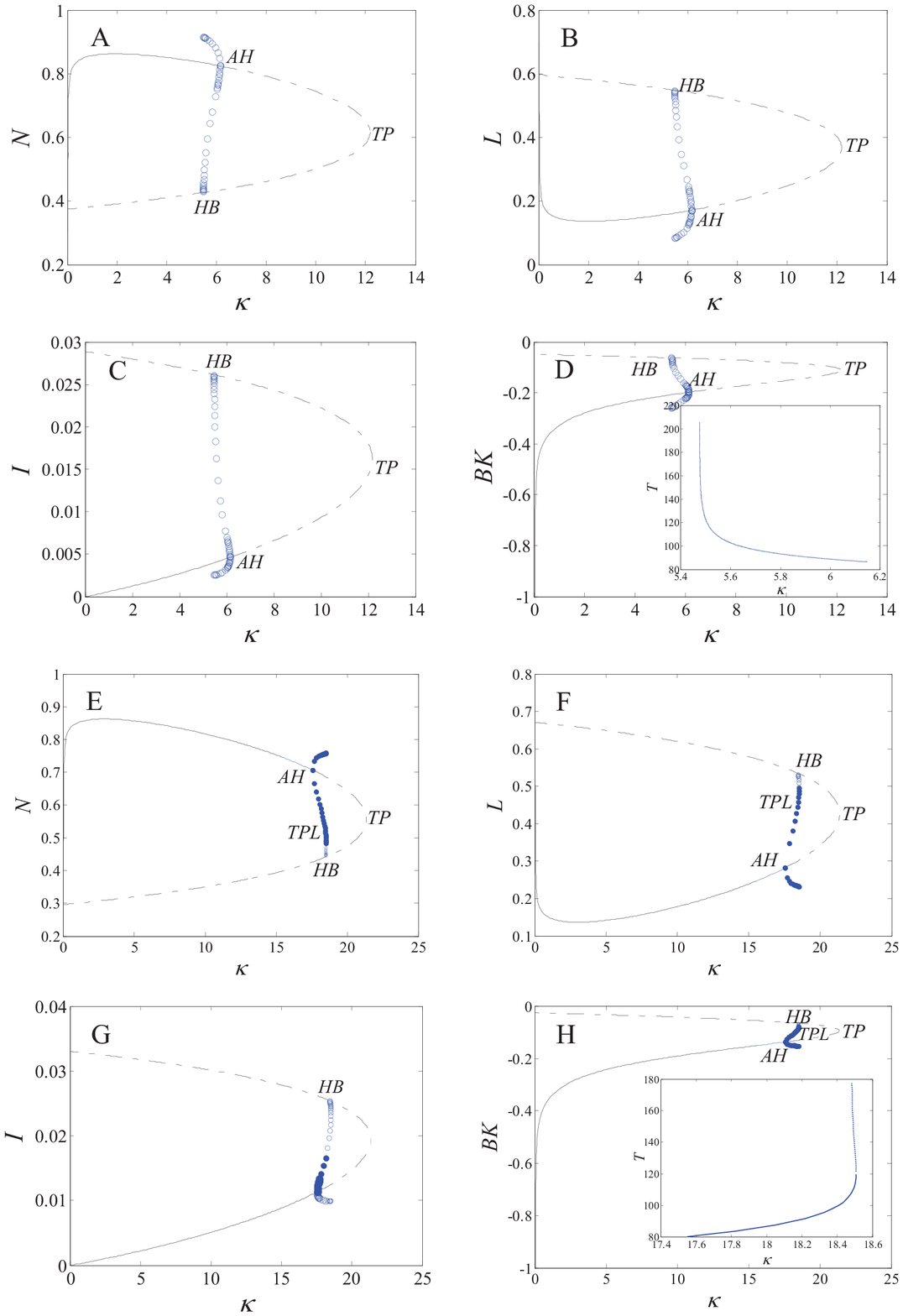}
\\
\caption{\textbf{One-dimensional bifurcation diagrams w.r.t. $\kappa$} for $\beta=0.14$ (\textbf{A-D}) and for $\beta=0.18$ (\textbf{E-H}). Solid lines correspond to stable equilibria and dashed lines to unstable equilibria; filled (open) circles correspond to maximum and minimum values of stable (unstable) oscillations. In $D$, $H$ the inset shows the bifurcation diagram of the period of oscillations w.r.t to $\kappa$. For $\beta=0.14$ there is a subcritical Andronov-Hopf bifurcation ($AH$) at $\kappa \approx 6.147$ ($(N, L, I, BK) = (0.824, 0.171, 0.005,-0.197)$). A stable lobby-free solution $(N, L, I, BK) = (0.333, 0, 0.667, 33.333)$ (not shown) also exists for all values of $\kappa$.\\
For  $\beta=0.18$ there is a supercritical Andronov-Hopf bifurcation ($AH$) at $\kappa \approx 17.55$ ($(N, L, I, BK) \approx (0.707, 0.281, 0.011,-0.137)$). The branch of  limit cycles has a turning point ($TPL$) at $\kappa \approx 18.508$. A stable lobby-free solution $(N, L, I, BK) = (0.333, 0, 0.667, 33.333)$ (not shown in Figure) also exists for all values of $\kappa$.}
\label{fig3}
\end{center}
\end{figure*}

To analyse the system's behaviour in the presence of control policy against lobbying, i.e. for $\kappa > 0$, we constructed the bifurcation diagrams w.r.t. $\kappa$. Figure \ref{fig3} depicts the one-dimensional bifurcation diagrams for $\beta=0.14$ (Figures \ref{fig3}A-D) and $\beta=0.18$ (Figures \ref{fig3}E-H). 

For $\beta=0.14$, a stable lobby-free solution $(N, L, I, BK) = (0.333, 0, 0.667, 33.333)$ exists for all values of the bifurcation parameter $\kappa$ (not shown in Figure \ref{fig3}A-D). For $\kappa < \approx 6.147$, the co-existence of all three types of academic behaviour is possible. This co-existence is characterized by high densities of neutrals (around 80\% of the total staff, see Figure \ref{fig3}A), moderate densities of lobbyists (around 19\% of the total staff, see Figure \ref{fig3}B), low densities of $I$ (around 1 \% of the total staff, see Figure \ref{fig3}C) and a relatively low/moderate potential for breakthroughs (around -0.3,see Figure \ref{fig3}D). At $\kappa \approx 6.147$ ($(N, L, I, BK) = (0.824, 0.171, 0.005,-0.197)$), the co-existence of equilibria looses its stability by a subcritical Andronov-Hopf bifurcation ($AH$) which marks the onset of a branch of unstable limit cycles. This branch of unstable limit cycles disappears at a homoclinic bifurcation ($HB$) at $\kappa \approx 5.475$ where the unstable limit cycle hits the saddle equilibrium. For $\kappa > 6.147$ the only stable solution is the lobby-free one characterized by a very high level of disciplinary diversity.   For even higher values of $\kappa$, a turning point ($TP$) appears at $\kappa \approx 12.1782$ ($(N, L, I, BK) = (0.616, 0.367, 0.017, -0.11)$), which however does not change the stability of solutions; on either side of the turning point, the equilibria are saddles.\\
For $\beta=0.18$, we get the bifurcation diagrams w.r.t. $\kappa$ depicted in Figure \ref{fig3}E-H. Now the subcritical Andronov-Hopf bifurcation ($AH$) turns into a supercritical at $\kappa \approx 17.55$ ($(N, L, I, BK) = (0.707, 0.281, 0.011,-0.137)$), giving birth to a branch of stable limit cycles. The stability of limit cycles is then lost through a turning point of limit cycles ($TPL$) at $\kappa \approx 18.508$. The branch of unstable limit cycles disappears through a homoclinic bifurcation ($HB$) at $\kappa \approx 18.48$ where the limit cycle hits the saddle equilibrium. For $\kappa > 18.508$ the only stable solution is the lobby-free state characterized by a very high level of potential for breakthroughs. For even higher values of $\kappa$, a turning point ($TP$) appears at $\kappa \approx 21.349$ ($(N, L, I, BK) = (0.554, 0.426, 0.02,-0.094)$) that does not change the stability of equilibria as on either side these are saddles.

\begin{figure*}[!ht]
\begin{center}
\includegraphics[scale=0.8]{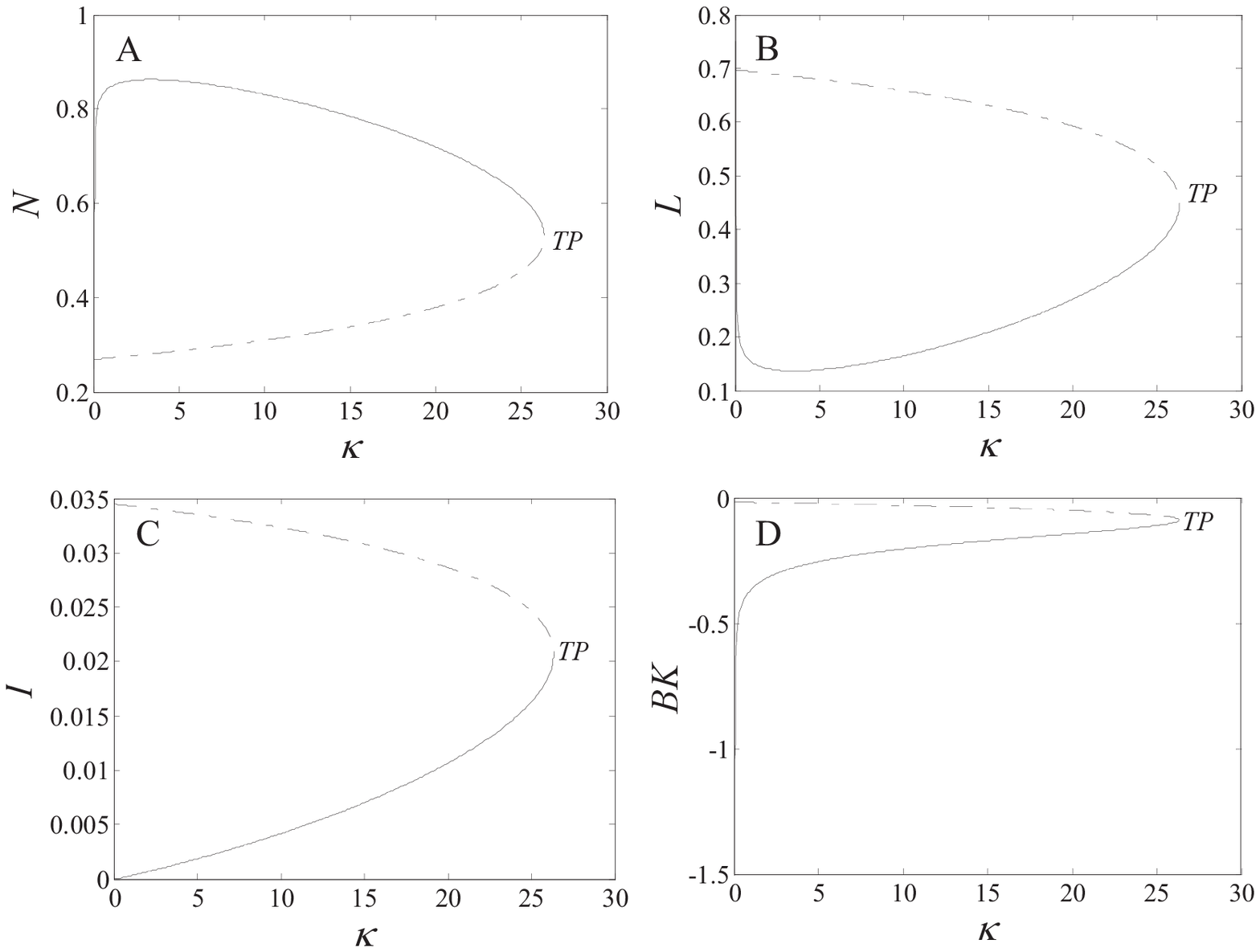}
\\
\caption{\textbf{One-dimensional bifurcation diagrams w.r.t. $\kappa$ for $\beta=0.2$.} \textbf{A}. $N$, \textbf{B}. $L$, \textbf{C}. $I$, \textbf{D}. $BK$. There is a turning point ($TP$) at $\kappa \approx 26.344$ ($(N,L,I,BK) \approx (0.53,0.45,0.02,-0.088)$). A stable lobby-free solution $(N, L, I, BK) = (0.333, 0, 0.667, 33.333)$ (not shown in the figure) also exists for all values of $\kappa$. Solid lines correspond to stable equilibria and dashed lines to unstable equilibria.}
\label{fig4}
\end{center}
\end{figure*}

By further increasing the value of $\beta$ to $\beta=0.2$ we get the bifurcation diagram w.r.t. $\kappa$ illustrated in Figure \ref{fig4}. The Andronov-Hopf bifurcation disappears and we end up with a single saddle-node bifurcation at $\kappa \approx 26.344$, $(N,L,I,BK) \approx (0.53,0.45,0.02,-0.088)$.

Taken all together, the above two one-dimensional bifurcation diagrams suggest the existence, in the two-dimensional parameter space ($\kappa$, $\beta$), of at least one Bautin (Generalized Andronov-Hopf) bifurcation marking the boundary between sub and supercritical Andronov-Hopf bifurcations, and potentially, of a Takens-Bogdanov bifurcation which marks the coincide of turning points with Andronov-Hopf bifurcations.

\begin{figure*}[!ht]
\begin{center}
\includegraphics[scale=0.95]{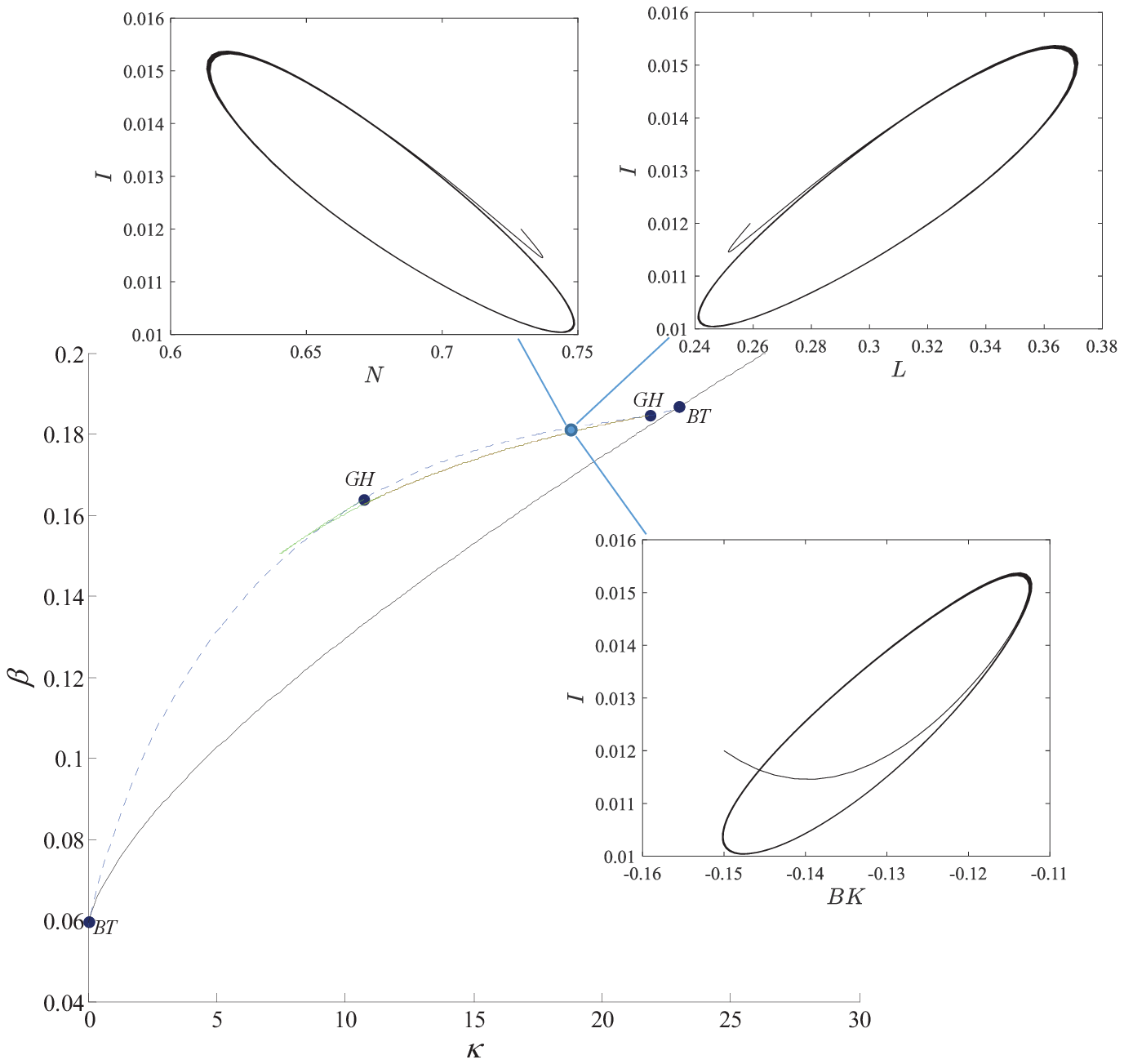}
\\
\caption{\textbf{Two-dimensional bifurcation diagram w.r.t. $\kappa$ and $\beta$.} There are two Bogdanov-Takens (BT) bifurcations and two Bautin (generalized Andronov-Hopf (GH)) bifurcations. The dotted (blue) line is the branch of Andronov-Hopf bifurcations. The branch of turning points of the limit cycles that connects the two GH bifurcations is shown with the solid (green) line; this branch exhibits a cusp point at $(\kappa$, $\beta)$ $\approx$ $(7.455,0.1503)$. The branch of the continuation of turning points passing through the two BT points is shown with the (black) solid line.
Phase portraits of sustained oscillations for $\kappa=18$, $\beta=0.18$ are also shown.}
\label{fig5}
\end{center}
\end{figure*}

To explore the overall system's behaviour in the two-dimensional parameter space ($\kappa$,$\beta$), we constructed the two-dimensional bifurcation diagram depicted in Figure \ref{fig5}. We also illustrate phase portraits of a sustained oscillation within the region of existence of stable limit cycles for $\kappa=18$, $\beta=0.18$ and initial conditions $N(0)=0.729$, $L(0)=0.259$, $I(0)=0.012$, $BK(0)=-0.15$.

As it is shown, there are two Bogdanov-Takens (BT) bifurcations, where the Andronov-Hopf bifurcations coincide with the turning points, at $(\kappa$, $\beta)$ $\approx$ $(0.045,0.059)$, $(N,L,I,BK) \approx (0.859,0.139,0.002,-0.253)$, and at $(\kappa$, $\beta)$ $\approx$ $(23.006,0.187)$, $(N,L,I,BK) \approx (0.546,0.434,0.02,-0.092)$. There are also two Bautin bifurcations (generalized Andronov-Hopf bifurcations (GH)) , which set the boundaries between sub and supercritical Andronov-Hopf bifurcations at $(\kappa$, $\beta)$ $\approx$ $(10.748,0.1639)$, $(N,L,I,d) \approx (0.788,0.205,0.007,-0.173)$, and  at $(\kappa$, $\beta)$ $\approx$ $(21.842,0.185)$, $(N,L,I,BK) \approx (0.607,0.376,0.017,-0.108)$. On the branch of the turning points of limit cycles that connects the two GH points (shown with (green) solid line), there is a cusp point at $(\kappa$, $\beta)$ $\approx$ $(7.455,0.1503)$. The region of stable sustained oscillations is bounded between the branch of Andronov-Hopf bifurcations connecting the two BT points (shown with dotted (blue) line), and the branch of turning points of branches of limit cycles (shown with (green) solid line).
At this point we should note that due to the test function that the MATCONT uses to perform continuation of Andronov-Hopf points, MATCONT reports also a branch of neutral saddles, that emerges from the BT bifurcation at $(\kappa$, $\beta)$ $\approx$ $(21.842,0.185)$. Neutral saddles satisfy also the test function of MATCONT for Andronov-Hopf points (as at these points the sum of two (non zero) real eigenvalues becomes zero), yet these points are not bifurcations. Hence, we don't show this branch in the two dimensional bifurcation diagram.

\section*{Discussion}

As well known in biology and ecology, diversity is a key factor for selection processes, system evolution \cite{Charlesworth} and ecosystem productivity \cite{Mazz}. The negative effects of excess of inbreeding are well known, producing worsening of genetic diversity and consequently lower competitive performance in the long term \cite{Hedrick}. In analogy to this, our model addresses a reciprocal negative feedback between the potential of knowledge breakthroughs and the growth of lobbies due to the related effects on disciplinary diversity.\\
An in-depth analysis of the Universities in USA \cite{Holl} demonstrated that the potential for scientific breakthroughs is significantly higher in relatively small and flexible research structures, whereas institutions characterized by high level of organization isomorphism, i.e. reduced disciplinary diversity in their research structures, show clear decline of their scientific innovation performances, due to the intrinsic tendency to work in established problems areas.\\
In Academia, the favoritism towards sub-standard researchers and/or relatives regardless of their scientific merit, has been long recognized. For example, it has been shown that nepotism results to significant negative correlation with scientific research performance  \cite{Allesina}. Nepotism is usually referred to the action that favors members of the own family regardless of their scientific merit, but here we claim that this should not be strictly limited to just phylogenetic relationships. Indeed, a broader cultural conception of familism/inbreeding can be described in which the ``family” corresponds to a scientific/cultural/behavioural category and the degree of kinship assessed by the level of affiliation/similarity to the same category (lobby).\\
In Europe, several studies have revealed high levels of academic inbreeding, where in some countries it reaches as much as 95\% \cite{Navaro,Soler}. On the other hand, the importance of intellectual and scientific diversity has been recognized in American universities, which in general do not hire their own PhD students. In USA, around 93\% of candidates to academic positions were reported as externals \cite{Blackwell,Pan}.
\\
In fact it has been reported that inbreeding inhibits the entrance of new and fresh ideas  as research partners within the same disciplines with strong ties over long periods may ``naturally" be entrapped to a clique \cite{Wonseok}. Interestingly, it has been argued that the production of quality outcomes may become limited not only by the individuals deficiencies, but also ``by cartels of mutually satisfied mediocrities” \cite{Gam_Ori}. Under such conditions, best individuals of small/weak groups are out competed by mediocres of strong large tribes. On the other hand, loose ties in such scientific cliques may provide opportunities/potential for integration of interdisciplinary knowledge from outsiders and therefore breakthroughs \cite{Wonseok}.

On the other hand, interdisciplinarity has been associated with research breakthroughs \cite{Holl}, but its quantitative and objective assessment is difficult \cite{Rafols}. Disciplinary diversity is a necessary condition for the growth and establishment of interdisciplinary research and can be inferred to some degree by ISI subject categories \cite{Porter}. However, disciplinary diversity is a necessary but not a sufficient condition to guarantee knowledge integration leading to truly interdisciplinary research; the latter rather depending on the successful interaction of ``local bodies" of knowledge \cite{Rafols}.\\

%For example, a study in the field of information systems showed an increase from 40\% to 80\% of co-authored articles compared to a drastic drop of single author works \cite{Wutchy}. 

Purposely, we did not model the ``size” of scientific production. Several studies have shown that research quality and breakthroughs are not directly related to  quantitative metrics such as journal impacts, number of publications and size of research teams \cite{Albert,Wonseok}. %This is reflected in all branches of science with major increase in the size of teams and in the number of research products but not in the number of breakthroughs \cite{Oh}.
Thus, we do not claim that lobbyists and neutrals necessary publish less or worse quality papers than independent researchers. Here, instead, we focused on the attitude of lobbyists as an opposite force to the disciplinary diversity, the necessary (yet not sufficient) condition for achieving breakthroughs in complex problems with important social impact. Since academic lobbies are obstacles to the advance of breakthrough knowledge, given their intrinsic resistance to dynamic changes of established equilibria \cite{Kerr}, the departments, in order to be successful in producing innovative research, have to mitigate the ``natural” trend of lobby formation. Breaking down dominant positions, despite the expected obvious internal opposition, will produce higher disciplinary diversity and by improving the cultural context, reinforce the long term performance of the academic structures. 
Thus, aiming to better scientific performances, one of the available options for policy actions is the promotion and increase of investments (both on recruitment and career advancement) in less represented fields.
But this is not so easy to achieve.
Moreover, where scientists are stressed to act according to ``publish or perish” orientation, in order to increase the number of publications and grants, open research questions are not efficiently faced \cite{Cao}.\\
 
\begin{figure*}[!ht]
\begin{center}
\includegraphics[scale=0.9]{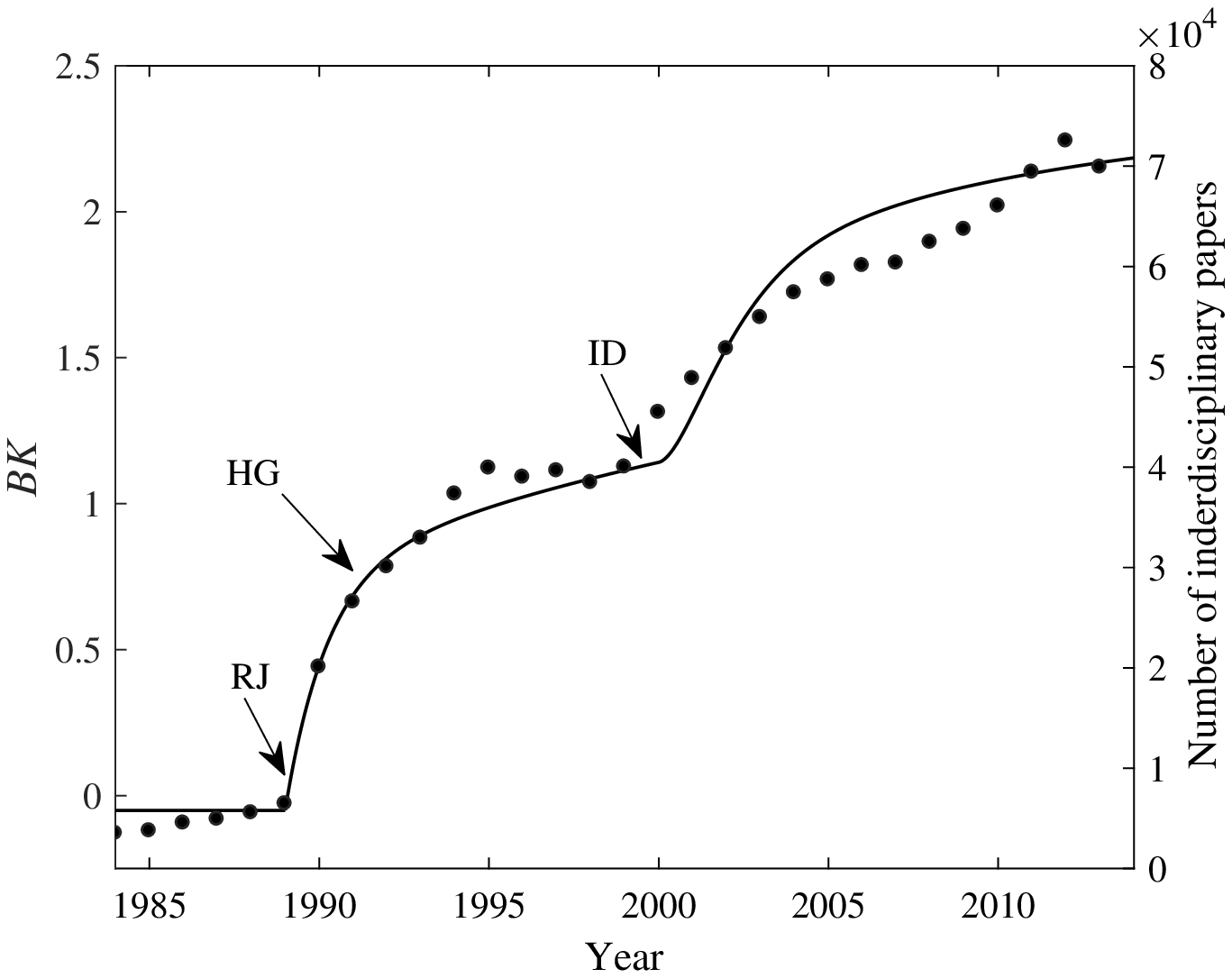}
\\
\caption{Phase transitions in disciplinary publications and related breakthrough knowledge. Dots correspond to number of interdisciplinary papers connecting mathematics and molecular biology resulting from the flow of ex-soviets scientists in the USA in 1990 (RJ), the raise in funding by the Human Genome Project in 1992 (HG) and internet diffusion in 2000 (ID) as reported in \cite{Phillips}. Solid line corresponds to model simulations with calibrated changes of the parametric values}.
\label{figRJ}
\end{center}
\end{figure*}

Building up on the above studies and conceptions, our mathematical model incorporated two categories of parameters, reflecting both internal and external factors. The internal factors reflect the direct and indirect interactions of individuals in an academic environment, while the external factor mirrors the control policy against lobbying, thus favoring the enhancement of disciplinary diversity. The model exhibits a complex dynamic behavior with multistability of states and sustained oscillations due to the interactions between the composition of the academic population and the cultural context related to the level of innovative thinking. 
These results are qualitatively similar to those reported for oscillating patterning of the commons in the so called coevolutionary game theory, with a coupled evolution of individual strategies and the related environmental context \cite{Weitz}.\\
Very clear examples of real data on the above discussed dynamics have been recently reported by Phillips \cite{Phillips}. A sudden and astonishing jump/phase transitions of both the number of publications and breakthroughs in mathematical and molecular biological sciences resulted by the increased flow of ex-Soviet scientists in the USA, after the dissolution of the Soviet Union starting in 1989. Moreover, other similar phase transitions, though to a lower extent to the above reported ``Russian jump" (RJ), were recorded in the period 1992–1996 with the \$3 billion funding of the Human Genome (HG) and, later on, in association with the Internet diffusion (ID) since 2000 \cite{Phillips}.
Our model is able to qualitatively represent these observations by a simple exploratory simulation exercise. (see Figure \ref{figRJ}. 

The RJ effect on bursting interdisciplinary research can be simulated by a sudden change of system diversity due to the entrance in the academic population of many new scientists with very different background. At the same time, the increase of resources due to HG funding can be reflected by policy intervention captured by increasing of $k$ parameter. On the other hand, the ID discontinuity started in 2000 can be obviously represented by a facilitated connection between research actors, i.e. a change in $\lambda$ and $\zeta$ parameters to reflect the faster exchange between neutral and independent individuals.\\

So, big sudden changes/phase transitions in the academic environment can be induced by external forces, i.e. control policies. In the presence of an external control action against lobbying and moderate rates of influence of lobbyists, our model predicts  critical points beyond which lobbying  disappears and there are again two stable states both characterized by zero levels of lobbying behaviour. One of them is characterized by the dominance of neutrals and low presence of independent researchers and thus moderate levels (potential) of research breakthroughs. The other one corresponds to very high levels/potential for research breakthroughs due to the dominance of independent researchers. \\
It should be noted that the intensity of external action that is necessary to moderate lobbying behaviour is relatively large with respect to the values of the other rates. This reflects the fact that (as also reported in \cite{BritishAcademy}) in order to tackle conservative attitudes, a high level of awareness and effort is required.
For higher values of the power of influence of lobbyists, the system behaviour is enriched with stable sustained oscillations in a relatively small window of the parameter space. These oscillations in the presence of a control action can be interpreted as steps back and forth dynamics between openness and closeness to new ideas as known to occur in societies with opposite tendency induced by concurrency \cite{Klapp}.

%\section*{Conclusions}
We should note that interdisciplinary knowledge does not imply superiority over disciplinary knowledge, nor that disciplinary research drives necessarily to the creation of lobbies. Advance of knowledge obviously occur in well established academic disciplines, however, in order to avoid decline in such positive performances, the research institutions need to be open to external development and recruitment \cite{Jacobs,Holl}. In our opinion, interdisciplinarity should not be regarded as an end in itself, but as a carrier of qualitative jumps/phase transitions of knowledge and technological advancements.
The proposed model demonstrates that under-performing academic systems become entrapped in autocatalytic lobby structures producing progressive reduction on the capability of knowledge advancement. On the other hand, high levels of innovation are related to systems characterized by efficient interdisciplinary network made probable by high levels of disciplinary diversity.
Intermediate conditions can create cycles of oscillatory performances, whereas poor quality can be changed only by external interventions injecting individuals from different disciplines into the system, thus destabilizing the natural lobby trends to increasing ``cognitive" rigidity. The entrance of fresh/novel ideas in a system most likely will produce jumps of scientific innovation. Then, ``good" science can be sustained by supporting research which is not limited to established main streams. This can be achieved only by introducing negative feedback on decision makers to force them away from academic inbreeding, thus reinforcing disciplinary diversity. \\

\section*{Materials and Methods}

%By definition, a lobby will put barriers and obstacles to the careers of independent researchers by influencing the recruitment process in a biased way to the advantage of its own growth.  However, we are not stating that it is intrinsically wrong to try to favor the own group, this will be the natural case also for an autonomous researcher. The main difference will be the level of hostility against the others. An independent researcher will appreciate a good active group, independently of its affiliation. Differently, a lobbyist will perceive as a risk the growth of a different group. These two opposite behaviours will objectively produce feedbacks on the diversity levels of the academic scenario with lobbyists and autonomous researchers decreasing and increasing, respectively, in the long term the rate of knowledge innovation.\\
A mean field model describing the dynamics of academic populations and related scientific potential to breakthroughs is presented. Three different types of academic staff individuals are identified and labeled as follows:

\begin{enumerate}
\item[$\bullet$] $L$ (Lobbyists which ``defend" their own discipline and group, hindering the entrance of new ideas and research and thus disciplinary diversity).
\item[$\bullet$]  $I$ (Independent researchers, reflecting disciplinary diversity and thus potential for advancing interdisciplinary research which may lead to breakthroughs).
\item[$\bullet$] $N$ (Neutrals who either enter into the Academia or they are already inside the system and are ``politically" passive in their preferences regarding the advance of diversity/interdisciplinarity).

\end{enumerate}
A fourth variable, $BK$, representing the level of interdisciplinarity (indicating the potential of achieving) breakthroughs. In our model, this is, by definition, enhanced by the presence of $I$ and inhibited by $L$. Negative (positive) values of $BK$ correspond to low (high) levels of such a potential. Negative and positive values of $BK$ could be defined in relation to an average level of scientific research performance, which is set to zero. Thus $BK$ is modeled through the following equation ($\dot{x}\equiv\frac{dx}{dt}$):

\begin{eqnarray}
\dot{BK}(t)=\theta \cdot I-\eta \cdot L -\epsilon \cdot BK
\label{eqn1}
\end{eqnarray}

where, $\theta$ is the rate of growth of $BK$ per independent researcher, $\eta$ is the rate of decline of $BK$ per lobbyist, and, $\epsilon$ is the  fade out rate of  $BK$  in the absence of any ``stimulus".\\
    
The model dynamics evolve according to the following rules:

\begin{enumerate}

\item[$\bullet$] Individuals leave the system (getting into pension or leaving the department) with a rate $\mu$. 

\item[$\bullet$] In an analogy to the infection rate in networked  epidemic models \cite{Brauer,Keeling}, a lobbyist may influence/``infect"/convince a neutral to become part of the lobby through direct interaction/contact. This is modeled through the transmission rate $\beta \equiv p \cdot s$, where $p$ is the per contact with a lobbyist probability that a neutral will be convinced to join lobbyists, and $s$ is the average number of contacts with lobbyists per unit time. Hence, $\beta$ is the average rate of contacts a neutral makes with lobbyists that are sufficient to make him/her to join lobbyists. Thus the mean field conversion rate of neutrals to lobbyists reads:

\begin{eqnarray}
r_{NL}=\beta \cdot N \cdot L.
\label{eqn2}
\end{eqnarray}

\item[$\bullet$] When $BK$ is positive, reflecting a relative high level of interdisciplinarity, thus a large potential for breakthroughs, the easier is for a neutral to become an independent researcher, and the easier is  to moderate/neutralize the behaviour of a lobbyist.

More specifically, neutrals become independent researchers with a rate $\lambda \cdot p_{NI}(BK) \cdot N$, where $p_{NI}(BK)$ is the logistic function:

\begin{eqnarray}
p_{NI}(BK)=\frac{1}{1+e^{-a (b \cdot BK-c)}}.
\label{eqn4}
\end{eqnarray}
\\

Figure \ref{figS2}A shows $p_{NI}(BK)$ for $a=1.5$, $b=10$, $c=5$. Note that there is nonzero small probability for a $N$ to behave/become as an $I$ even for (small) negative values of $BK$.

\begin{figure*}[!ht]
\begin{center}
\includegraphics[scale=0.9]{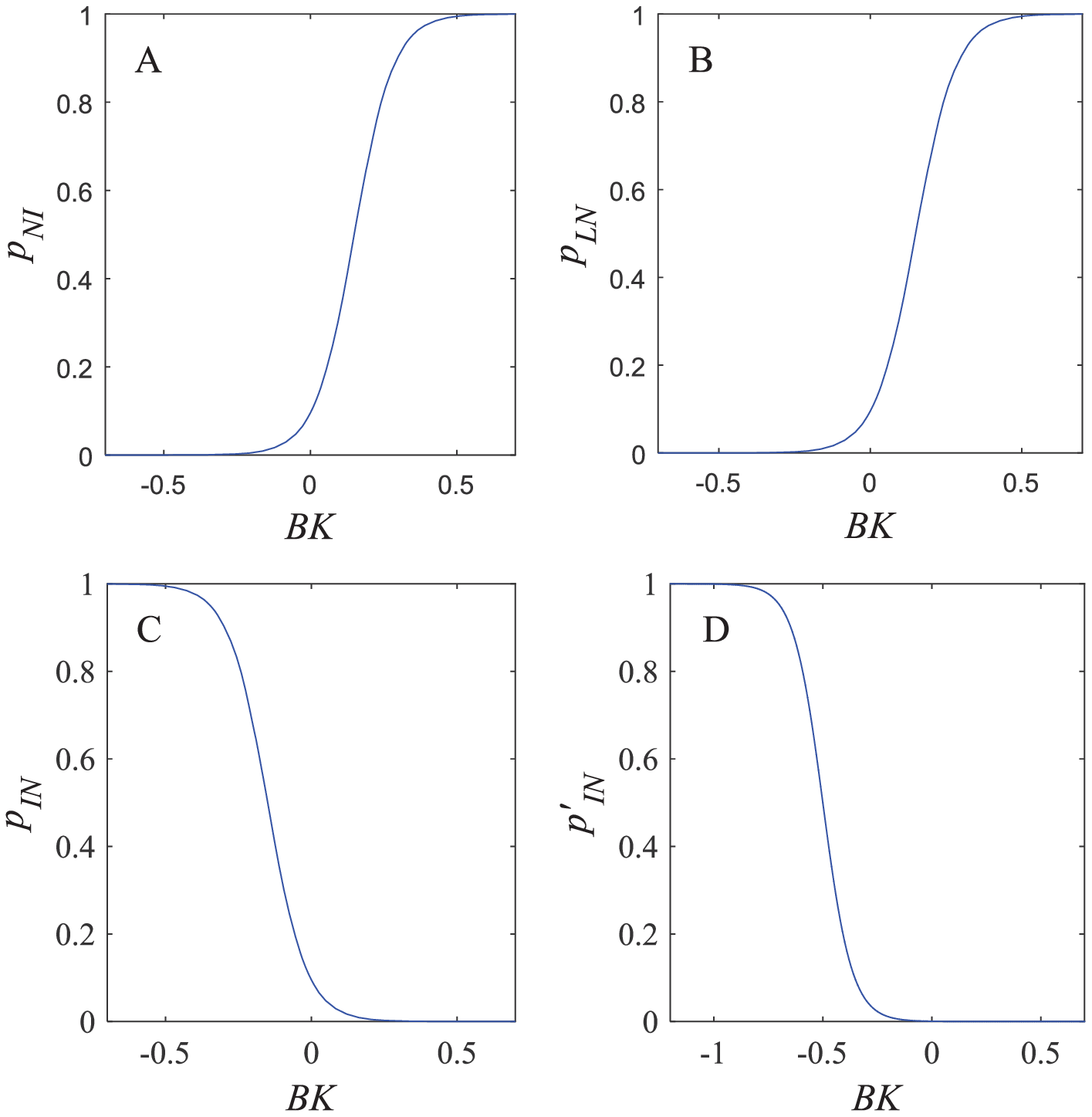}
\\
\caption{Logistic functions as scaling factors to the rates of transformation w.r.t. $BK$ \textbf{A}. $p_{NI}(BK)$ scales the rate at which a neutral individual becomes an independent researcher;$a=1.5$, $b=10$, $c=-1.5$). \textbf{B}. $p_{LN}(BK)$ scales the rate at which a lobbyist becomes neutral. \textbf{C}.  $p_{IN}(BK)$ scales the rate at which an independent researcher becomes neutral ($a=1.5$, $b=10$, $c=-1.5$). \textbf{D}.  $p'_{LN}(BK)$ scales the rate at which a lobbyist individual becomes neutral by the implementation of a control policy. $a=1.5$, $b=10$, $c=-1.5$.}
\label{figS2}
\end{center}
\end{figure*}

\item[$\bullet$] In a similar manner, lobbyists are neutralized with a rate $r_{LN}(BK)=\gamma \cdot p_{LN}(BK) \cdot L$, where $p_{LN}(BK)$ is the logistic function, given by:

\begin{eqnarray}
p_{LN}(BK)=\frac{1}{1+e^{-a (b \cdot BK-c)}}.
\label{eqn6}
\end{eqnarray}
\\
Figure \ref{figS2}B shows $p_{LN}(BK)$ for $a=1.5$, $b=10$, $c=5$. Again, there is nonzero small rate at which  $L$ are neutralized even
for (small) negative values of $BK$.

%Eq.~(\ref{pdeqn10})%

\item[$\bullet$] When $BK$ is negative, reflecting a relative low level of  interdisciplinarity/ potential for breakthroughs, the easier is for an independent researcher to be neutralized.

More specifically, independent researchers leave the system/neutralize with a rate $\zeta \cdot p_{IN}(BK)\cdot I$, where $p_{IN}(BK)$ is the logistic function:

\begin{eqnarray}
p_{IN}(BK)=1-\frac{1}{1+e^{-a (b \cdot BK-c)}}
\label{eqn8}.
\end{eqnarray}
\\
Figure \ref{figS2}C shows $p_{IN}(BK)$ for $a=1.5$, $b=10$, $c=-1.5$. Note, that there is a nonzero small rate at which $I$ are neutralized for small positive values of $BK$.

\item[$\bullet$] At low levels of interdisciplinarity, as reflected by negative values of $BK$, we also impose a ``control policy" that exerts an external action/feedback aiming to ``moderate the conservative forces” i.e. to neutralize lobbying behaviour in the presence of relatively low levels of interdisciplinarity. 

Accordingly, lobbyists are neutralized with a rate $\kappa \cdot p'_{LN}(BK) \cdot L$, where $p'_{LN}(BK)$ is the logistic function:

\begin{eqnarray}
p'_{LN}(BK)=1-\frac{1}{1+e^{-a (b \cdot BK-c)}}.
\label{eqn10}
\end{eqnarray}
\\

Hence, $\kappa \cdot p'_{LN}(BK)$ is the rate at which the external policy neutralizes lobbying. Its inverse can be regarded as the mean latent period, i.e. the period between the application of the policy and neutralization.

Figure \ref{figS2}D shows $p'_{LN}(BK)$ for $a=1.5$, $b=10$, $c=-1.5$. Hence the external control action is activated only for relatively large negative values of $BK$.

\end{enumerate}

Our choice to model the rate of transformations of the indirect interaction (i.e. the transformations w.r.t. $BK$) by the logistic function is based not only on well established biological/population dynamics theoretical concepts and experimental studies. Recent social dynamics studies including experimental data have shown that the logistic function models also the impact of social interconnected relationships (e.g.cooperation, friendship, communication, influence, consensus formation and decision making) \cite{Christakis,Xing}.

In summary, based on the above assumptions, the mean field model reads:

\begin{equation}
\begin{split}
\dot{N}= \mu \cdot (1-N) + \kappa \cdot p'_{LN}(BK) \cdot L + \zeta \cdot p_{IN}(BK) \cdot I \\
-\lambda \cdot p_{NI}(BK) \cdot N-\beta \cdot N \cdot L+\gamma \cdot p_{LN}(BK) \cdot L
\end{split}
\label{eqn11}
\end{equation}
\\
\begin{equation}
\dot{L}=-\mu \cdot L+\beta \cdot N \cdot L - \kappa \cdot p'_{LN}(BK) \cdot L -\gamma \cdot p_{LN}(BK) \cdot L
\label{eqn12}
\end{equation}
\\
\begin{equation}
\dot{I}=-\mu \cdot I +\lambda \cdot p_{NI}(BK) \cdot N- \zeta \cdot p_{IN}(BK) \cdot I
\label{eqn13}
\end{equation}
\\
\begin{equation}
\dot{BK}(t)=\theta \cdot I-\eta \cdot L -\epsilon \cdot BK
\label{eqn14}
\end{equation}
\\
Note that if the initial conditions are chosen so that $N+L+I=1$, the above system preserves mass as $\dot{N}+\dot{L}+\dot{I}=0$. Hence Eq.~(\ref{eqn13}) can be omitted for a numerical analysis point of view.
Computations were performed using MATCONT ~\cite{matcont1}. The continuation of solutions past critical points was performed using the Moore-Penrose continuation ~\cite{matcont1} and the absolute and relative error for the Newton-Raphson iterations were set equal to $1.E-06$. The tolerance of test functions used to detect criticalities was set equal to $1.E-05$. The computations of limit cycles was performed using 20 mesh points and 4 collocation points.\\
A summary of the model parameters and variables are shown in Table 1.

\begin{table}[ht!]
  \centering
  \caption{Model parameters and variables.}
  \label{tab:table1}
  \begin{tabular}{p{2cm}p{7.5cm}p{3cm}}
    \toprule
    Symbol & Definition & Units\\
    \midrule
    $N$ & density of neutrals & dimensionless\\
    $L$ & density of lobbyists & dimensionless\\
    $I$ & density of independent researchers & dimensionless\\
    $\mu$ & probability at which individuals get into pension, and/or leaving their department& $years^{-1}$\\
    $BK$ & potential for breakthroughs in knowledge  & dimensionless\\
     $\beta$ & average rate of contacts a neutral makes with lobbyists that are sufficient to make him/her to join lobbyists& $contacts \cdot years^{-1}$\\
    $\zeta \cdot p_{IN}(BK)$ & rate at which an independent researcher becomes neutral; $p_{IN}(BK)$ is the logistic function scaling the rate w.r.t. $BK$& $years^{-1}$\\
    $\lambda \cdot p_{NI}(BK)$ & rate at which a neutral becomes independent researcher; $p_{NI}(BK)$ is the logistic function scaling the rate w.r.t. $BK$& $years^{-1}$\\
    $\gamma \cdot p_{LN}(BK)$ & rate at which a lobbyist becomes neutral; $p_{LN}(BK)$ is the logistic function scaling the rate w.r.t. $BK$& $years^{-1}$\\
    $\kappa \cdot p'_{LN}(BK)$ & rate at which external policy neutralizes lobbying. It's inverse can be regarded as the mean latent period, i.e. the period between the implementation of the policy and neutralization; $p'_{LN}(BK)$ is the logistic function scaling the rate w.r.t. $BK$& $years^{-1}$ \\
    $\theta$ & rate of growth of $BK$ per independent researcher& $years^{-1}$\\
    $\eta$ & rate of decline of $BK$ per lobbyist& $years^{-1}$\\
    $\epsilon$ & rate of decline of $BK$ per neutral& $years^{-1}$\\
    \bottomrule
  \end{tabular}
\end{table}

\nolinenumbers

\section*{Author Contributions}
S.M. conceptual model design. C.S. and L.R. mathematical model development and numerical analysis. F.G. and G.T. simulations and numerical analysis. All authors contribute to manuscript preparation.

%\section*{Competing Financial Interests}
%There are no Competing Financial Interests.

%\newpage 

%\section*{Figures and Tables}

%\newpage 

%\newpage 

%\newpage 

%\newpage 

%\newpage 

%\newpage 

%\section*{Supplementary Materials}

%\beginsupplement

%Eq.~(\ref{pdeqn10})%

%\begin{figure*}[!ht]
%\begin{center}
%\includegraphics[scale=0.52]{Figure3tot.eps}

%Fig.~\ref{fig3} 

\end{document}